\documentclass[letterpaper]{article} 
\usepackage[]{aaai24}  
\usepackage{times}  
\usepackage{helvet}  
\usepackage{courier}  
\usepackage[hyphens]{url}  
\usepackage{graphicx} 
\usepackage{caption} 
\usepackage{algorithm}
\usepackage{algorithmic}

\usepackage{booktabs} 
\usepackage{multirow}
\usepackage{bbding}

\usepackage{newfloat}
\usepackage{listings}
\DeclareCaptionStyle{ruled}{labelfont=normalfont,labelsep=colon,strut=off} 
\floatstyle{ruled}
\newfloat{listing}{tb}{lst}{}
\floatname{listing}{Listing}
\usepackage{amsmath}

\title{Joint Edge Optimization Deep Unfolding Network \\ for Accelerated MRI Reconstruction}
\author{
    Yue Cai,
    Yu Luo,
    Jie Ling,
    Shun Yao
}
\usepackage{bibentry}

\begin{document}

\maketitle

\begin{abstract}
Magnetic Resonance Imaging (MRI) is a widely used imaging technique, however it has the limitation of long scanning time. Though previous model-based and learning-based  MRI reconstruction methods have shown promising performance, most of them have not fully utilized the edge prior of MR images, and there is still much room for improvement. 
In this paper, we build a joint edge optimization model that not only incorporates individual regularizers specific to both the MR image and the edges, but also enforces a co-regularizer to effectively establish a stronger correlation between them. Specifically, the edge information is defined through a non-edge probability map to guide the image reconstruction during the optimization process. Meanwhile, the regularizers pertaining to images and edges are incorporated into a deep unfolding network to automatically learn their respective inherent a-priori information.
Numerical experiments, consisting of  multi-coil and single-coil MRI data with different sampling schemes at a variety of sampling factors, demonstrate that the proposed method outperforms other state-of-the-art methods.  

\end{abstract}

\section{Introduction}
\label{sec:introduction}
\begin{figure}[t]
\centering
\includegraphics[width=0.9\columnwidth]{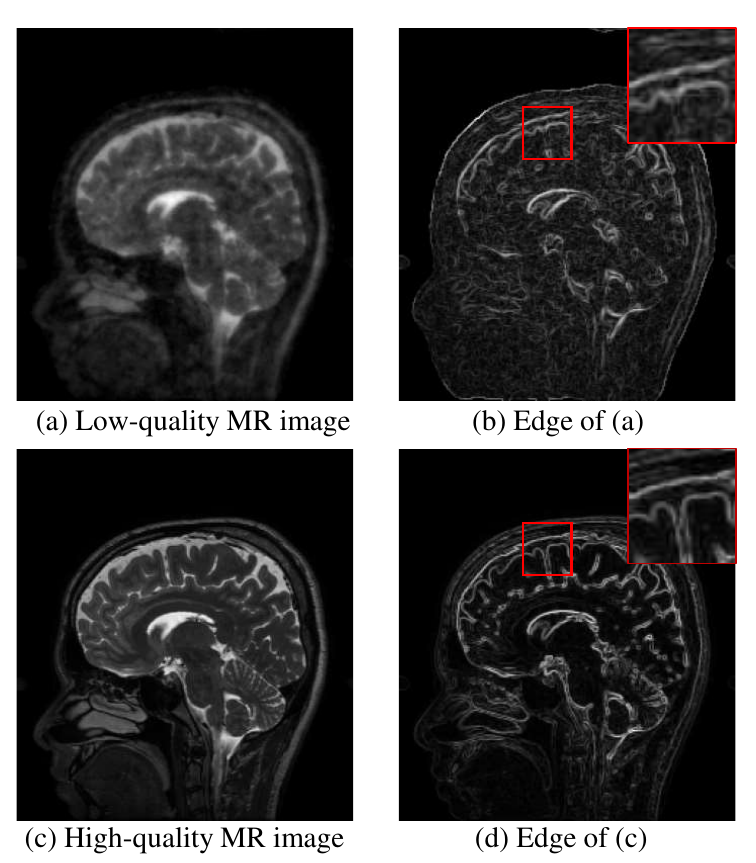} 
\caption{(a) The low-quality MR image; (b) The edges of (a); (c) The high-quality MR image; (d) The edges of (c).}
\label{fig_intro}
\end{figure}
Magnetic resonance imaging (MRI) is a widely used imaging technique in the medical field. However, it has the limitation of a long scanning time.
To accelerate the scanning, on one side, the raw data in \emph{k}-space are partially sampled and various signal processing methods (such as compressed sensing) are proposed to reconstruct high-quality MR image from the undersampled data. On the other side, parallel imaging technique that utilizes multiple coils to scan the anatomy simultaneously is developed. Therefore, the typical parallel MRI reconstruction problem with classical compressed sensing can be modeled as follows:
\begin{equation}
\mathop {\arg \min }\limits_x \sum\limits_{i = 1}^{{n}} {\left\| {UF{S_i}x - {y_i}} \right\|_2^2 + \lambda _1 \Phi(x)}, 
\label{eq:1}
\end{equation}
where $n$ is the number of coils, $x \in {C^M}$ is the MR image to be reconstructed,
${S_i} \in {C^M}$ is the $i^{th}$ coil sensitivity estimation map, $F$ is the Fourier transform operator, $U \in {C^{N \times M}}(M >> N)$ is the undersampled matrix, ${y_i} \in {C^N}$ is the partially sampled \emph{k}-space data, 
$\lambda _1$ is a trade-off parameter, and $\Phi(x)$ is the regularization term that usually enforces sparsity.

In the past years, various regularization terms with corresponding optimization algorithms were proposed in the MRI reconstruction problem and achieved good performance. 
Among these model-based methods, 
most of them focus on searching the suitable transform to prompt sparsity (e.g. total variation \cite{rudin1992nonlinear}, spline wavelet transforms \cite{shen2010wavelet}, adaptive learned dictionary \cite{ravishankar2010mr}
and combined transforms \cite{luo2020cosparse})  
or elaborately designing the variants of sparsity regularizers (e.g. $l_1$ regularizer \cite{ning2013magnetic}, $l_q$ regularizer \cite{wang2013recovery}, and group sparsity \cite{chen2014exploiting}).
In addition to sparsity, researchers are also exploring some other priors of MR images such as low rank \cite{yao2018efficient}, non-local similarity \cite{qu2014magnetic}, etc.
Nevertheless, these methods constrain the entire image with a unified prior indiscriminately and ignore the diverse structures inside the image, which may result in blurred edges or some unwanted artifacts, such as the fake edges in smooth regions.
Additionally, edges are important in images as they dominate the content of an image. As shown in Figure \ref{fig_intro}, a high-quality image is typically accompanied by a distinct and sharp edge, whereas a low-quality image is often characterized by a blurred or indistinct edge. Better preservation of edges is crucial for better image reconstruction.

Considering the above issues, some works have been proposed to impose extra structure or edge priors to reduce the artifacts and improve the quality of the final reconstructed image. Methods with edge priors for MRI reconstruction can be roughly classified into two categories: the traditional model-based methods and the deep learning-based methods. The first category incorporates regularization of the edge into the optimization model, e.g. adaptive edge-preserving regularization \cite{belge2000wavelet} and geometrically structured approximation \cite{ji2016image}. This kind of methods have two issues, the first is the introduce of edge prior term will increase the complexity of the numerical algorithm or even cause the non-convergence issue. The second issue is that the design of edge prior is critical, a less adequate assumption could be counterproductive. Though the deep learning based methods do not need to face the above dilemma, they have a challenge about how to effectively utilize the edge prior. For most deep learning based methods with edge prior, as far as we know, edges are used as the guidance by being concatenated with the input in a direct \cite{nazeri2019edgeconnect, luthra2021eformer, chen2021edbgan} or a fancy way  \cite{yang2023fast}. The focus is either on the extraction of the edges or the fusion mechanism between the edge and the MR image. However, how the concatenation play the role of guidance is unclear and is often ignored. The potential of edge has not been fully exploited.

In order to effectively utilize the edge prior as well as avoid the interference of inaccurate prior assumption, the proposed MRI reconstruction model needs to have the following two properties. First, to avoid the influence of inaccurate assumption, the regularizations of the MR image to be recovered do not need to have a explicit expression like in traditional model based methods. Second, to fully exploit the potential of the edge information, effective correlations between edges and the target MR image should be established. Ground on the above discussions, in this paper, we propose a joint edge optimization parallel MRI reconstruction model that seeks a tighter coupling between edges and images to obtain a high-quality reconstructed MR image.  A co-regularizer between the MR image and the edge-related variables is utilized to link the two closely, so as to achieve a better reconstruction during the iterate optimization with the assistance of their respective constraints. Specifically, the edge-related variables are characterized as a probability map that measures the probability of a region being a non-edge (smooth region). Meanwhile, to reduce reliance on hand-crafted priors, the regularizations regarding the target MR image and the probability map are incorporated into the proximal mapping operator simulated by neural network modules. The deep unfolding schema is utilized to automatically learn the inherent a-priori information of the MR image and non-edge probability map  in the optimization process.
Overall, the main contributions of our proposed method are as follows:

\begin{itemize}
\item 
Unlike previous deep learning-based approaches that simply concatenate the edge with MR images to incorporate edge priors, our proposed joint edge optimization MRI reconstruction model employs a more intimate integration way to fuse the edge and images for mutually guidance and enhancement. 
\item 
Unlike traditional model-based approaches that rely on manually designed regularizations, our proposed model automatically learns inherent priors of images and edges through the network with a deep unfolding schema.
\item In our experiments, datasets from different imaging scenarios are utilized for validation. The quantization metrics are superior compared to other state-of-the-art methods under different acceleration factor and different sampling schemes.
\end{itemize}

\section{Related work}

\subsection{Accelerated parallel imaging}
In accelerated parallel imaging, the algorithm aims to reconstruct high-quality MR images from partially sampled multi-coil \emph{k}-space data.
The methods of MRI reconstruction in parallel imaging scenarios can be categorized into traditional model-based and data-driven methods. In traditional methods, GRAPPA \cite{Griswold_Jakob_Heidemann_Nittka_Jellus_Wang_Kiefer_Haase_2002} uses interpolation in \emph{k}-space; SENSE \cite{pruessmann1999sense} exploits sensitivity estimation maps in the spatial domain to eliminate aliasing artifacts. In addition to this, some of the methods solve the optimization model (\ref{eq:1}) based on the theory of compressed sensing in parallel imaging scenarios, such as ESPIRiT \cite{Uecker_Lai_Murphy_Virtue_Elad_Pauly_Vasanawala_Lustig_2014}, which combines a multi-coil degradation mechanism and a constrained regularity term for the reconstruction. However, the reconstructed image quality of traditional methods is not stable and the imaging speed is slow. 

With the development of deep learning techniques, artificial neural networks have achieved excellence in many tasks in the field of computer vision \cite{lecun2015deep,luo2023local,luo2023pseudo}. Meanwhile, the data-driven methods appear and the researchers explore the potential of neural networks in the field of medical imaging. In the accelerated MRI reconstruction task, Wang et al. \cite{wang2016accelerating} and Schlemper et al. \cite{schlemper2017deep} proposed for the early time to use convolutional neural networks to learn the mapping relationship between zero-filled undersampled images to high-quality reconstructed images. Hyun et al.\cite{Hyun_Kim_Lee_Lee_Seo_2018} utilized U-Net \cite{ronneberger2015u} which is an efficient encoder-decoder architecture in segmentation tasks, as the reconstruction modeling framework. Similar works have been done by Sun et al. \cite{sun2018compressed} utilizing recursive dilated networks. Some methods \cite{quan2018compressed}\cite{yang2017dagan} restore zero-filled degraded images based on the generative adversarial network architecture as a means of removing aliasing artifacts and redundant noise. In addition, since the original MRI data is in complex form, researchers \cite{wang2020deepcomplexmri} applied complex convolution operation to jointly process the real and imaginary parts of the MR image to preserve the magnitude and phase information of the image. Moreover, Considering the importance of edge information restoration, EAMRI \cite{yang2023fast} utilizes the edge detector to extract the image edge, which is used as guidance for constructing the attention map. However, the "off-the-shelf" edge detectors have their own limitations which will migrate to the reconstruction algorithm. 
Despite their reconstruction result is generally better than traditional methods with lower inference time, the data-driven reconstruction methods described above are not conducive to the clinical practice due to the lack of interpretability resulting from the black-box nature of neural networks.

\subsection{Advances in Deep Unfolding for MRI reconstruction}
The innovative deep unfolding methods are initially introduced to sparse coding approximations \cite{gregor2010learning}, fusing optimization-driven iterative algorithms with the expressive power of neural networks. This methodology has seen successful application across a variety of low-level image processing tasks \cite{monga2021algorithm}.

For instance, the Denoising Convolutional Neural Network (DPDNN) \cite{dong2018denoising} and the Model-based Deep Learning architecture (MoDL) \cite{aggarwal2018modl} utilize deep unfolding schema by employing neural networks to embody model-based regularization terms within image domain. E2EVarNet \cite{sriram2020end} integrate a gradient descent optimization for \emph{k}-space data enhancements in MRI reconstruction, unfolding the first-order derivatives of regularization function with the neural network. 
HQS-Net \cite{xin2022learned} uses a half quadratic splitting method to process the reconstruction model and integrates the neural network module into the optimization process to design an efficient and lightweight reconstruction method; HUMUS-Net \cite{fabian2022humus} integrates the multi-scale network structure and the attention mechanism into the depth-expanded reconstruction method to achieve high-quality reconstruction results, but due to the large size of the parameters of the network model, it is not that efficient in training. 
Meanwhile, methods like ISTA-Net \cite{zhang2018ista} and ADMM-CSNET \cite{yang2018admm} take a leap forward by unfolding iterative soft-thresholding and alternating direction multiplier algorithms, respectively, into learning-based models that combine optimized algorithm with data-driven component.
VS-Net \cite{duan2019vs} unfolding the iterative refinement optimization process with neural network to constructing a lightweight architecture, while MGDUN \cite{yang2022model} uses neural network structures to semantically replace traditional sampling and modality conversion processes, particularly highlighting its prowess in tasks such as cross-modality super-resolution.

 A noteworthy aspect of deep unfolding schema is its inherent reduction of hand-craft design, leaning towards an autonomous optimization. Moreover, the deployment of deep unfolding has not only augmented the clarity and interpretability of computational models but also enhanced the precision of the eventual solutions. 

\section{Method}
\subsection{Model}
The signal intensity of MR images typically demonstrates a piecewise smooth distribution, which is attributed to the correlation with the underlying tissue structure. With this piecewise smooth property, a redundant system can provide a sparse representation of the MR image, generating coefficients with large amplitude along the locations of image edges and coefficients with near-zero (or zero) amplitude within the smooth regions (non-edge regions). Based on this, the edges and the image can be effectively associated with each other. In this paper, we introduce a probability map named $P_{ne}$ to quantify the probability of a region being smooth (non-edge) and utilize it to constrain the values of transformation coefficient inside the smooth regions to be close to zero. In addition, as MR images depict the structural information of anatomical tissues, they have certain geometrical structures. Consequently, the smooth regions within these MR images are not randomly distributed but conform to specific patterns and characteristics. Therefore, the probability map $P_{ne}$ that measures the likelihood of a region being smooth also should fulfill certain properties and can be imposed with regularizations to satisfy the corresponding properties.

Based on the aforementioned discussion, our proposed joint edge optimization MRI reconstruction model can be formulated as follows:

\begin{equation}
\begin{aligned}
E(x,{P_{ne}}) &= \frac{1}{2}\sum\limits_{i = 1}^{{n}} {\left\| {UF{S_i}x - {y_i}} \right\|_2^2} + {\lambda _1}\Phi(x)\\
&+ \frac{\rho }{2}\left\| {{P_{ne}}*Wx} \right\|_2^2 + {\lambda _2}\Psi({P_{ne}}),
\end{aligned}
\label{eq:2}
\end{equation}
where $*$ denotes a pixel-wise product\footnote{$*$ is omitted in the following text for brevity.}, ${P_{ne}}$ is the introduced non-edge probability map related to the edge information. The value of ${P_{ne}}$ ranges from $0$ to $1$. When its value is $0$, it indicates that the likelihood of the pixel belonging to a non-edge is $0$, and we do not constrain the corresponding transformation coefficient. On the other hand, when the value is $1$, it suggests that there is a high probability that this pixel belongs to a smooth (non-edge) region, and we expect the transformation coefficient to be close to $0$. Conversely, when the images are updated, the value of ${P_{ne}}$ will be decreased  or increased accordingly based on whether the corresponding pixel belongs to edge or not. $W$ denotes the stationary Haar wavelet transformation. ${\lambda _1}$, ${\lambda _2}$, and $\rho $ are the trade-off parameters of different regularization terms.

The optimization model (\ref{eq:2}) consists of four terms. The first is the fidelity term, which ensures the data consistency between the reconstructed MR image $x$ and the measured \emph{k}-space data. The third term is the co-regularizer between image $x$ and edge-related variables ${P_{ne}}$, linked by a transformation $W$, aiming to promote the piece-wise smoothness property of image $x$.  The second term $\Phi(x)$  and the forth term $\Psi({P_{ne}})$ are the regularizers of the image $x$ and the probability map ${P_{ne}}$ respectively. Instead of imposing hand-crafted priors, we learn the inherent priors of images and edges automatically through the network. 

\begin{figure*}[t]
\centering
\includegraphics[width=0.96\textwidth]{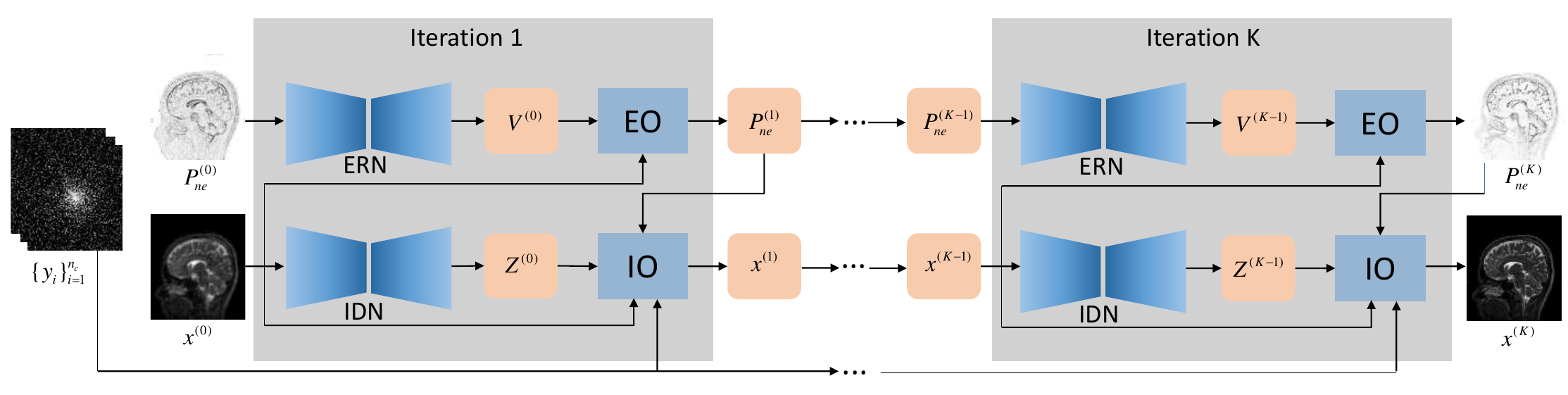} 
\caption{The overall iterative framework of the joint edge optimization deep unfolding network for accelerated MRI}
\label{fig2}
\end{figure*}

\subsection{Algorithms}
In Model (\ref{eq:2}), two variables are included, i.e., the reconstructed image $x$ and the non-edge probability map $P_{ne}$. To solve this optimal model, we decouple the two variables  $\{x, P_{ne}\}$ and optimize them alternatively by solving the following subproblem (\ref{eq:3}) and (\ref{eq:4}):
\begin{equation}
P_{ne}^{(k + 1)} = \mathop {\arg \min }\limits_{{P_{ne}}} \frac{\rho }{2}\left\| {{P_{ne}}W{x^{(k)}}} \right\|_2^2 + {\lambda _2}\Psi({P_{ne}}),
\label{eq:3}
\end{equation}

\begin{equation}
\begin{aligned}
    {x^{(k + 1)}} &= \mathop {\arg \min }\limits_x \frac{1}{2}\sum\limits_{i = 1}^{{n}} {\left\| {UF{S_i}x - {y_i}} \right\|_2^2} \\
    &+ {\lambda _1}\Phi(x) + \frac{\rho }{2}\left\| {P_{ne}^{(k + 1)}Wx} \right\|_2^2,
\end{aligned}
\label{eq:4}
\end{equation}
where $k$ denotes the stage index of the iteration. 
As seen in the above subproblems, the co-regularizer between the reconstructed image and the non-edge probability map exists in both two subproblems. Hence in the alternating optimization process of two variables, they will guide each other for better recovery.
For the solution of the two subproblems, we will introduce them in detail in the following.

\subsubsection{Edge Information Solver}
Inspired by the Half Quadratic Splitting (HQS) \cite{geman1995nonlinear}
method, the subproblem (\ref{eq:3}) is re-formulated into the following form as:
\begin{equation}
\mathop {\min }\limits_{{P_{ne}},V} \frac{\rho }{2}\left\| {{P_{ne}}W{x^{(k)}}} \right\|_2^2 + \frac{\alpha }{2}\left\| {V - {P_{ne}}} \right\|_2^2 + {\lambda _2}\Psi(V),
\label{eq:5}
\end{equation}
where $V$ is the introduced auxiliary variable, $\alpha$ is the trade-off parameter.

Based the above optimization problem, we form the optimization problem for the auxiliary variable V by extracting the terms related to V as follows:
\begin{equation}
{V^{(k)}} = \mathop {\arg \min }\limits_V \frac{\alpha }{2}\left\| {V - P_{ne}^{(k)}} \right\|_2^2 + {\lambda _2}\Psi(V).
\label{eq:6}
\end{equation}
For the optimization of auxiliary variables $V$, it is actually the proximal mapping of
regularizer $\Psi$. 
We use the efficient proximal gradient descent (PGD) to solve the problem as follow:
\begin{equation}
{V^{(k)}} = prox_{\frac{\lambda_2 }{\alpha}\Psi}(P_{ne}^{(k)}).
\label{eq:8}
\end{equation}

After updating $V^{(k)}$, the non-edge probability map $P_{ne}$ can be solved by the following optimization model:
\begin{equation}
P_{ne}^{(k + 1)} = \mathop {\arg \min }\limits_{{P_{ne}}} \frac{\rho }{2}\left\| {{P_{ne}}W{x^{(k)}}} \right\|_2^2 + \frac{\alpha }{2}\left\| {{V^{(k)}} - {P_{ne}}} \right\|_2^2,
\label{eq:7}
\end{equation}
which has a closed-form solution as follow:
\begin{equation}
P_{ne}^{(k + 1)} = \alpha {V^{(k)}}/[\rho {(W{x^{(k)}})^2} + \alpha],
\label{eq:9}
\end{equation}
where $/$ denotes element-wise division.

\subsubsection{Reconstructed Image Solver}
Similar to the solution of the non-edge probability map $P_{ne}$, we also solve subproblem (\ref{eq:4}) based on the HQS by introducing the auxiliary variable $Z$ for the reconstructed image $x$. Subproblem (\ref{eq:4}) can be rewritten as follow:
\begin{equation}
\begin{aligned}
  \mathop {\min }\limits_{x,Z} \frac{1}{2}\sum\limits_{i = 1}^{{n}} &{\left\| {UF{S_i}x - {y_i}} \right\|_2^2}  + \frac{\rho }{2}\left\| {P_{ne}^{(k + 1)}Wx} \right\|_2^2 \\
  &+ \frac{\beta }{2}\left\| {Z - x} \right\|_2^2 + {\lambda _1}\Phi(Z),
\end{aligned}
\label{eq:10}
\end{equation}
where $\beta$ is the trade-off parameter. 

Similarly, based the above optimization problem, we construct the optimization problem for the auxiliary variable Z by extracting the terms related to $Z$ as follows:
\begin{equation}
{Z^{(k)}} = \mathop {\arg \min }\limits_Z \frac{\beta }{2}\left\| {Z - {x^{(k)}}} \right\|_2^2 + {\lambda _1}\Phi(Z),
\label{eq:11}
\end{equation}
which is also a proximal mapping of regularizer $\Phi$, denoted as:
\begin{equation}
{Z^{(k)}} = prox_{\frac{\lambda_1 }{\beta}\Phi}({x^{(k)}}).
\label{eq:13}
\end{equation}

Subsequently, we extract the $x$ related terms to build the optimization model for $x$ as follows:
\begin{equation}
\begin{aligned}
    &{x^{(k + 1)}} = \mathop {\arg \min }\limits_x \frac{1}{2}\sum\limits_{i = 1}^{{n}} {\left\| {UF{S_i}x - {y_i}} \right\|_2^2} \\
    &+ \frac{\rho }{2}\left\| {P_{ne}^{(k + 1)}Wx} \right\|_2^2 + \frac{\beta }{2}\left\| {{Z^{(k)}} - x} \right\|_2^2.
\end{aligned}
\label{eq:12}
\end{equation}
As to the optimization of image $x$, it can be obtained by the classical conjugate gradient algorithm. However as the iterative solution process involves repeated undersampling operator $U$, Fourier transform $F$ and wavelet transform $W$, the computational burden is an unavoidable issue. Inspired by the effectiveness of the optimization model using lightweight single-step gradient descent in DPDNN \cite{dong2018denoising}, we adopt the single-step gradient descent for approximating the reconstructed image $x$.
The iterative schema for updating the reconstructed image $x$ is shown below:
\begin{equation}
\begin{aligned}
    &{x^{(k + 1)}} = {x^{(k)}} - s[\sum\limits_{i = 1}^{{n}} {S_i^H{F^H}{U^H}(UF{S_i}{x^{(k)}} - {y_i})} \\
    &+ \rho {W^H}P_{ne}^{(k + 1)H}(P_{ne}^{(k + 1)}W{x^{(k)}}) - \beta ({Z^{(k)}} - {x^{(k)}})],
\end{aligned}
\label{eq:14}
\end{equation}
where $s$ is the step size of the single-step gradient descent.
$H$ denotes the corresponding conjugate operation.

\subsection{Deep Unfolding Network}
The principle of deep unfolding is that the modules of optimize-based approach which need to be manually designed are difficult to determine, so the deep unfolding schema utilize the neural networks to replaced the undetermined module or operator, introducing the representing capability of deep neural networks into the optimizing process.  In traditional model-based methods, the design of the regularization terms $\Phi$ and $\Psi$ are crucial for the final reconstruction. In this paper, to avoid the reliance on hand-crafted prior and the interference of inaccurate assumptions, the neural networks are utilized to automatically learn the inherent priors. Specifically,  
since the regularization terms $\Phi$ and $\Psi$ only occur in the proximal operator in Eq. (\ref{eq:8}) and Eq. (\ref{eq:13}), we consider the substitution of neural networks for the proximal operator.
Thus for Eq. (\ref{eq:8})  we utilize an edge recovery network (\emph{ERN}) to replace the proximal gradient operator $prox_{\Psi}$. 
And for Eq. (\ref{eq:13}), the proximal gradient operator $prox_{\Phi}$ is substituted using image denoising network (\emph{IDN}).  
Therefore, the optimization for the auxiliary variables $V$ and $Z$ can be rewritten in the following form:
\begin{equation}
{V^{(k)}} = ERN(P_{ne}^{(k)},{\theta_e ^{(k)}}),
\label{eq:15}
\end{equation}
\begin{equation}
{Z^{(k)}} = IDN({x^{(k)}},{\theta_i ^{(k)}}),
\label{eq:16}
\end{equation}
where ${\theta_e ^{(k)}}$ and ${\theta_i ^{(k)}}$ are the parameters of the edge recovery network \emph{ERN} and the image denoising network \emph{IDN} at the ${k}^{th}$ iteration, respectively. As to the specific network structure design of \emph{ERN} and \emph{IDN}, we utilize the standard U-Net architecture \cite{ronneberger2015u} which is suitable for the image-to-image conversion task due to its encoder-decoder structure that extracts features from images at different scales through multi-layer up-sampling and down-sampling operations. Moreover, the standard U-Net architecture is combined with the residual connection from the module input to output in our method so as to avoid gradient vanishing issue \cite{he2016deep}. 

\begin{figure*}[t]
\centering
\includegraphics[width=1.0\textwidth]{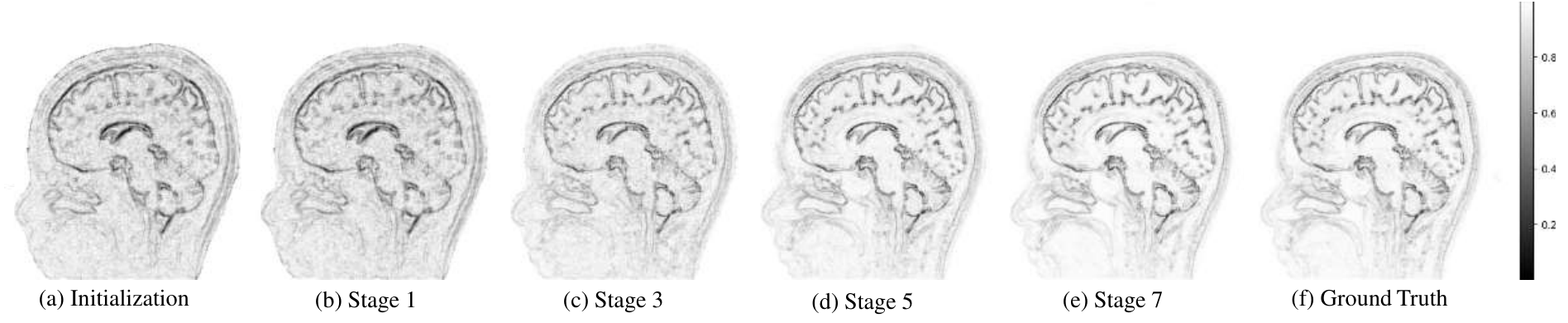} 
\caption{Non-edge probability map visualization. (a) and (f) is the initialization and ground truth of non-edge probability map respectively. From (b) to (e), we visualize the non-edge probability map of increasing stages.}
\label{nep_evolution}
\end{figure*}

In the following, we present an overall reconstruction framework for our joint edge optimization model. The optimization process consists of $K$ iterative stages, each containing four optimization steps. First, the edge auxiliary variable $V$ is optimized using the edge recovery network (\emph{ERN}). Then, the non-edge probability map ${P_{ne}}$ is optimized based on the closed-form solution of Eq. (\ref{eq:9}), known as edge optimization (\emph{EO}). The reconstructed image auxiliary variable $Z$ is optimized using the image denoising network (\emph{IDN}), and finally, the reconstructed image $x$ is optimized using single-step gradient descent by Eq. (\ref{eq:14}), known as image optimization (\emph{IO}). The detailed optimizing steps at each stage are described in Algorithm~\ref{alg0}. As for the initialization of the image ${x^{(0)}}$, we process the partially sampled \emph{k}-space data of the multi-coils using the zero-filled method to obtain the image domain data and later fuse the multi-coils data into one channel using sensitivity estimation maps, i.e., ${x^{(0)}} = \sum\limits_{i = 1}^{{n}} {S_i^H{F^H}{y_i}}$; 
As for the initialization of the non-edge probability map $P_{ne}^{(0)}$, first the stationary haar wavelet transform operator $W$ transforms the initial MR image to the wavelet domain, then the coefficients of the high-frequency channel are normalized to [0,1] with a maximum-minimum normalization operator $N$. Finally, the initial $P_{ne}^0$ is obtained by inverting the normalized coefficients as follows:
\begin{equation}
P_{ne}^{(0)} = 1 - N(|W{x^{(0)}}|),
\label{eq:P_ne0}
\end{equation}
where $|W{x^{(0)}}|$ is the coefficients of the high-frequency channel by pixel-wise absolute operation.
The iterative reconstruction framework is shown in Figure \ref{fig2}, where $P_{ne}^{(0)}$ and ${x^{(0)}}$ are used as inputs to the $K$-stage deep unfolding reconstruction network to obtain the optimized edge information $P_{ne}^{(K)}$ and the final reconstructed MR image ${x^{(K)}}$. In Figure \ref{nep_evolution}, we visualize the non-edge probability map while showing the evolution procedure at different iteration stages.

\begin{algorithm}[tb] 
	\renewcommand{\algorithmicrequire}{\textbf{Input:}}
	\renewcommand{\algorithmicensure}{\textbf{Output:}}
	\caption{Alternating iteration procedure} 
	\label{alg0} 
	\begin{algorithmic}[1]
		\REQUIRE undersampling zero-filled multi-coil \emph{k}-space data $\{ {y_i}\} _{i = 1}^{{n}}$;
		sensitivity maps $\{ {S_i}\} _{i = 1}^{{n}}$;
		undersampling matrix $U$;
		hyperparameters $\alpha$, $\rho$, $\beta$, $s$.	
		\ENSURE reconstructed MR image $x^{(K)}$. 
		\STATE Initialization:\\
  ${x^{(0)}} = \sum\limits_{i = 1}^{{n}} {S_i^H{F^H}{y_i}}$; \\$P_{ne}^{(0)} = 1 - N(|W{x^{(0)}}|)$.
		\FOR{$k=1,2,\ldots,K$}
		\STATE{$V^{(k)} = ERN(P_{ne}^{(k)})$};
		\STATE{$P_{ne}^{(k + 1)} = \alpha {V^{(k)}}/[\rho {(W{x^{(k)}})^2} + \alpha ]$};
		\STATE{$Z^{(k)} = IDN({x^{(k)}})$};
		\STATE{${x^{(k + 1)}} = {x^{(k)}} - s[\sum\limits_{i = 1}^{{n}} {S_i^H{F^H}{U^H}(UF{S_i}{x^{(k)}} - {y_i})}  +$ \\
			$\rho {W^H}P_{ne}^{(k + 1)H}(P_{ne}^{(k + 1)}W{x^{(k)}}) - \beta ({Z^{(k)}} - {x^{(k)}})]$};
		\ENDFOR
	\end{algorithmic}
\end{algorithm}

\subsection{Training Strategy}

Subsequent to the construction of the deep unfolding iterative framework, the training of our reconstruction model follows an end-to-end supervised learning paradigm. Similar with the process of initialization, fully sampled \emph{k}-space data are exploited to create ground truth image $\hat x$ and the non-edge probability map $\hat P_{ne}$. The training strategy utilizes a multi-component loss function, composed of the mean square error (MSE) for fidelity in image reconstruction and the L1 norm for precision in edge information recovery. Consequently, the total loss function $L$ for the joint edge optimization within the reconstruction model is formulated as follows:

\begin{equation}
L = \sum\limits_{i = 0} ^ {N_s} {{\gamma _1}\left\| {x_i^{K} - {{\hat x}_i}} \right\|_2^2 + {\gamma _2}{{\left\| P^K_{ne_i} - {{\hat P_{ne_i}}} \right\|}_1}}.
\label{eq:17}
\end{equation}

Here, $N_s$ symbolizes the number of samples in the training dataset sample scale while ${\gamma _1}$ and ${\gamma _2}$ are the respective trade-off parameters that modulate the loss components' contribution to $L$.
In addition, adjusting all of the undetermined parameters in the iterative process manually is a cumbersome task accompanied by a degree of uncertainty. 
Therefore, to elevate the model training efficiency, all of the undetermined parameters along with the parameters from the edge recovery network (ERN) and the image denoising network (IDN), denoted by $\{{\theta_e ^{(k)}}\} _{k = 1}^K$ and $\{ {\theta_i ^{(k)}}\} _{k = 1}^K$, are collected as trainable entities. This framework ensures the self-adjusting capability of the hyper-parameters throughout the back-propagation process, fostering their inherent adaptability within the reconstruction architecture.
With regards to multi-stage iterative model configurations, our investigative results utilizes a non-shared parameter strategy. Such an approach amplifies the quality of the reconstructed image in the context of our joint edge information optimization framework. Accordingly, the learnable parameter set for the proposed reconstruction model is expressed as $\Theta  = \{ {\theta_e ^{(k)}},{\theta_i ^{(k)}},{\rho ^{(k)}},{\alpha ^{(k)}},{\beta ^{(k)}},{s^{(k)}}\} _{k = 1}^K$, encompassing all the iterations and facilitating a tailored optimization at each discrete stage of the unfolding process.

\begin{table}[t]
\center
\begin{tabular}{c|c|c|c|c}
\hline
\multicolumn{1}{c|}{\multirow{2}{*}{Method}} & \multicolumn{2}{c|}{Cartesian} & \multicolumn{2}{c}{Random}\\ \cline{2-5}
 & R=6  & R=10 & R=6 & R=10 \\ 
 \hline
 Zero-Filled& 28.19 & 26.66 & 28.26 & 26.70  \\
 U-Net& 37.78 & 35.34 & 38.23 & 35.11  \\
 DCCNN& 39.05 & 37.22 & 39.29 & 37.33 \\ 
 MoDL& 39.56 & 37.33  & 39.24 & 37.35  \\ 
 VS-Net& 39.81 & 37.54 & 40.12 & 37.48 \\
 RecurrentVarNet& 39.66 & 37.41 & 39.74 & 37.52 \\
 EAMRI& 39.76 & 37.55 & 39.43 & 37.49 \\
 ours& \textbf{39.92} & \textbf{37.61} & \textbf{40.15} & \textbf{37.60} \\
 \hline
\end{tabular}
\caption{The PSNR results in multi-coil with different kinds of accelerated factor and sampling pattern.}
\label{table_multi}
\end{table}

\section{Experimental Evaluation}

\subsection{Implementation Details}
For our experiments, we employed two NVIDIA RTX 3090 Ti GPUs for our hardware and Pytorch is used as the deep learning framework for building the training and inferring pipeline.
The training process utilized the ADAM optimizer \cite{kingma2014adam}, which is known for its adaptive learning capabilities. An initial learning rate of 0.01 was chosen to ensure a strong starting point for optimization, and we applied a cosine decay learning rate schedule to reduce the learning rate smoothly over epochs for enhancing the model's convergence ability.
Training was conducted with a small batch size of 2. We repeatedly passed the dataset through the model for 180 epochs to ensure thorough learning and convergence.
Our model architecture chosen K = 7 stages in total, based on empirical evidence suggesting this to be optimal for the performance of our method. Each stage's parameters operated under a non-shared strategy, which provided the flexibility needed to adapt to each specific set of the optimized stages.
The ERN and IDN networks were built upon the original U-Net architecture, capitalizing on its symmetrical design with max-pooling layers in the encoder for undersampling and an equal number of transconvolution layers in the decoder for upsampling, ensuring detailed feature extraction and restoration.
We opted for Kaiming initialization \cite{he2015delving} for setting up the neural network parameters, a method designed to maintain adaptive variance as a countermeasure against gradients diminishing too rapidly during training. Regarding the loss function, we integrated trade-off parameters ${\gamma _1}$ and ${\gamma _2}$, weighted as 1 and 0.1 to balance the total loss function.

\begin{table}[t]
\center
\begin{tabular}{c|c|c|c|c}
\hline
\multicolumn{1}{c|}{\multirow{2}{*}{Method}} & \multicolumn{2}{c|}{Cartesian Random} & \multicolumn{2}{c}{Cartesian Equidistant}\\ \cline{2-5}
 & R=6  & R=10 & R=6 & R=10 \\ 
 \hline
 Zero-Filled& 28.46 & 26.87 & 28.30 & 26.74  \\ 
 U-Net& 30.94 & 28.52  & 30.87 & 28.44  \\
 DCCNN& 31.20 & 28.66  & 31.21 & 28.59  \\ 
 RefineGAN& 31.28 & 28.73  & 31.29 & 28.69  \\ 
 ADMM-CSNet& 31.16 & 28.70  & 31.22 & 28.65  \\ 
 HQS-Net& 31.42 & 28.84  & 31.33 & 28.73  \\ 
 EAMRI& 31.45 & 28.80  & 31.37 & 28.74 \\ 
 ours& \textbf{31.52} & \textbf{28.93} & \textbf{31.44} & \textbf{28.87} \\ 
 \hline
\end{tabular}
\caption{The PSNR results in single-coil with different kinds of accelerated factor and sampling pattern.}
\label{table_single}
\end{table}

\begin{figure*}[t]
\centering
\includegraphics[width=1.0\textwidth]{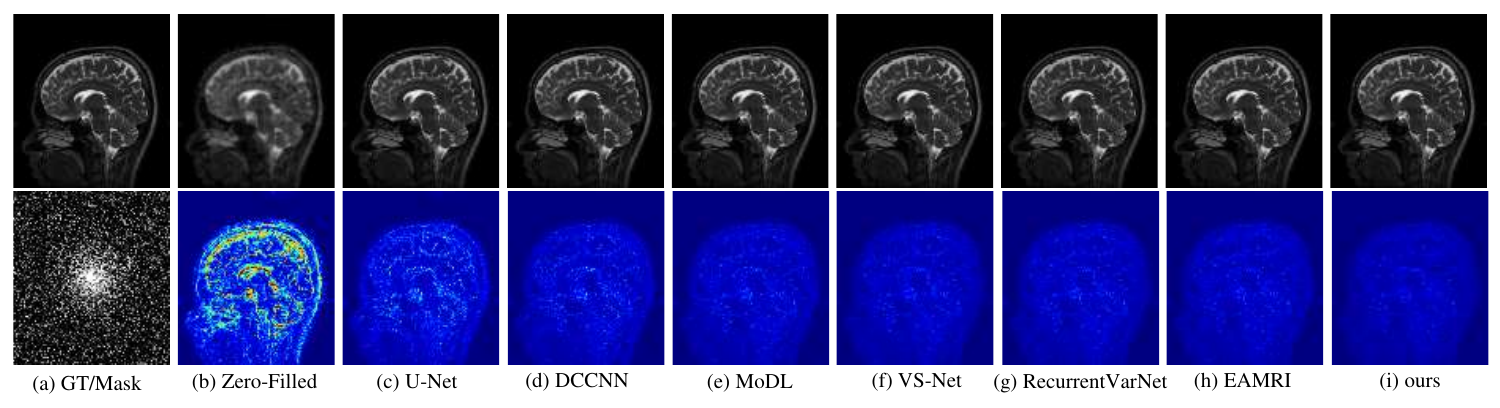} 
\caption{Results on multi-coil dataset using 6x random sampling. (a) Ground truth and random mask with the accelerated factor of 6. (b) The zero-filled reconstruction result and the error map; (c)-(i) The reconstructed images and the error maps by U-Net\cite{Hyun_Kim_Lee_Lee_Seo_2018}, DCCNN\cite{schlemper2017deep}, MoDL\cite{aggarwal2018modl}, VS-Net\cite{duan2019vs}, RecurrentVarNet\cite{yiasemis2022recurrent}, EAMRI\cite{yang2023fast} and our proposed method respectively.}
\label{fig_multi}
\end{figure*}

\begin{figure*}[t]
\centering
\includegraphics[width=1.0\textwidth]{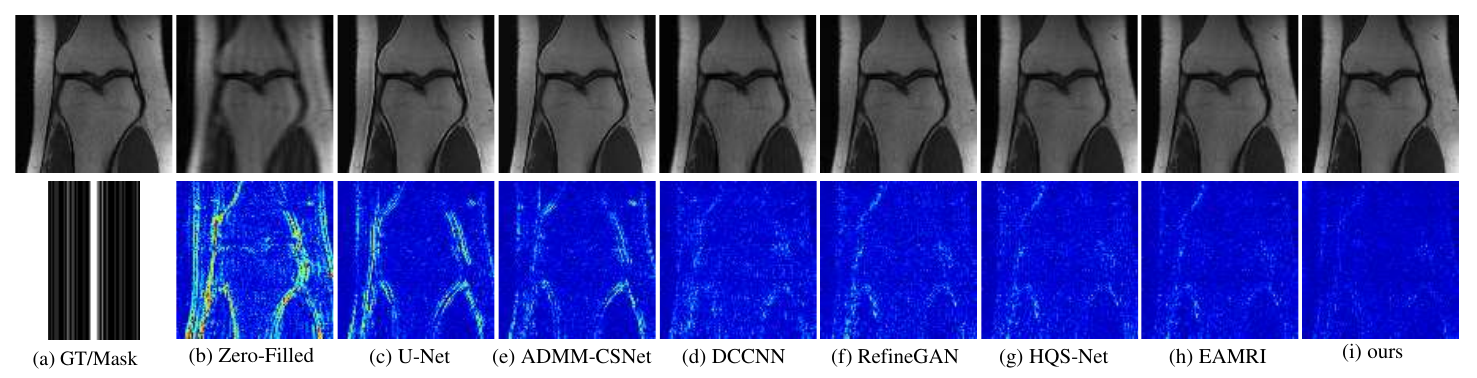} 
\caption{Results on single-coil dataset using 6x random Cartesian sampling. (a) Ground truth and random Cartesian mask with the accelerated factor of 6. (b) The zero-filled reconstruction result and the error map; (c)-(i) The reconstructed images and the error maps by U-Net\cite{Hyun_Kim_Lee_Lee_Seo_2018}, ADMM-CSNet\cite{yang2018admm}, DCCNN\cite{schlemper2017deep}, RefineGAN\cite{quan2018compressed}, HQS-Net\cite{xin2022learned}, EAMRI\cite{yang2023fast} and our proposed method respectively.}
\label{fig_single}
\end{figure*}

\subsection{Dataset}

For a comprehensive assessment, we engage in comparative experiments using both multi-coil and single-coil MRI data. Our ablation study and parameter strategy analysis are conducted on the multi-coil dataset for the parallel imaging scenario.

The multi-coil dataset employed is sourced from the dataset made public in conjunction with the MoDL study \cite{aggarwal2018modl}. This dataset is derived from a 3D T2 CUBE sequence captured using a 12-coil parallel imaging setup. It encompasses a total of 360 slices allocated for training, complemented by a validation set consisting of 164 slices. Each image slice adheres to a resolution of 256x232. The sensitivity maps, which are indispensable for various parallel imaging reconstruction techniques, were calculated using the ESPIRIT algorithm \cite{Uecker_Lai_Murphy_Virtue_Elad_Pauly_Vasanawala_Lustig_2014} that utilize the center region of \emph{k}-space to reconstruct the MR image, aiming at considering the most informative part of the MRI signal for image reconstruction.

In addition to multi-coil data, our methodology was also validated on the fastMRI single-coil knee dataset \cite{Zbontar_Knoll_Sriram_Muckley_Bruno_Defazio_Parente_Geras_Katsnelson_Chandarana_etal_2018} including varied contrasts such as proton-density weighting with fat-saturated (PDFS) and proton-density (PD) imaging modalities.
We randomly sampled 20\% of both training and testing samples from the aforementioned repository.
The single-coil dataset setting in our experiment included 4034 slices for training purposes and 745 slices designated for testing. Adhering to common evaluation protocols, all slices were uniformly cropped centrally to a standardized dimension of 320x320 pixels.

\subsection{Quantitative and Qualitative Evaluation}

To comprehensively assess the effectiveness of our proposed methodology, we conduct evaluations using both multi-coil and single-coil MRI datasets. The Peak Signal-to-Noise Ratio (PSNR) is utilized as the principal metric for quantitative analysis.
In total, we have conducted comparisons with nine additional methods. Among them, U-Net\cite{Hyun_Kim_Lee_Lee_Seo_2018}, DCCNN\cite{schlemper2017deep} and RefineGAN\cite{quan2018compressed} use a pure data-driven paradigm to reconstruct MR images in the image domain using neural network models. RefineGAN uses generative adversarial network and bi-domain cyclic loss function. MoDL\cite{aggarwal2018modl}, VS-Net\cite{duan2019vs}, RecurrentVarNet\cite{yiasemis2022recurrent}, ADMM-CSNet\cite{yang2018admm} and HQS-Net\cite{xin2022learned} use deep unfolding paradigm, which utilize neural networks for unfolding the optimization process of MR images. EAMRI\cite{yang2023fast} takes the image edge information into account in the reconstruction method and they extract the edge information based on the edge detection operator to guide the generation of the attention map. All competing methods are fine-tuned to their optimal settings for a fair comparison. 

In the context of multi-coil MRI data, we explore scenarios involving Cartesian and random undersampling schemes at acceleration factors of 6x and 10x, as documented in Table \ref{table_multi}. The experiment results indicate that regardless of the undersampling templates or acceleration rates, our method outperforms the other techniques in terms of PSNR values. This is further supported by error maps presented in Figure \ref{fig_multi}, which clearly demonstrate that the proposed approach achieves the minimum reconstruction error.
For single-coil MRI data, experiments are imparted at both 6x and 10x acceleration using random and equidistant Cartesian undersampling schemes. The quantitative results presented in Table \ref{table_single} reveal our method's superior performance in PSNR metric outpacing other algorithms. Supplementing our quantitative analysis is a visual examination of the single-coil reconstruction outcomes, shown in Figure \ref{fig_single}, which showcase that our method not only produces reconstructions with precise edge details but also exhibits minimal errors when compared to the fully-sampled reference. Overall, these visual results illustrate  our method's adaptability and performance across variously accelerated MRI scenarios on both multi-coil and single-coil configurations.

\subsection{Ablation Study}
\subsubsection{Effectiveness of Joint Edge Optimization}

\begin{figure}[t]
\centering
\includegraphics[width=0.9\columnwidth]{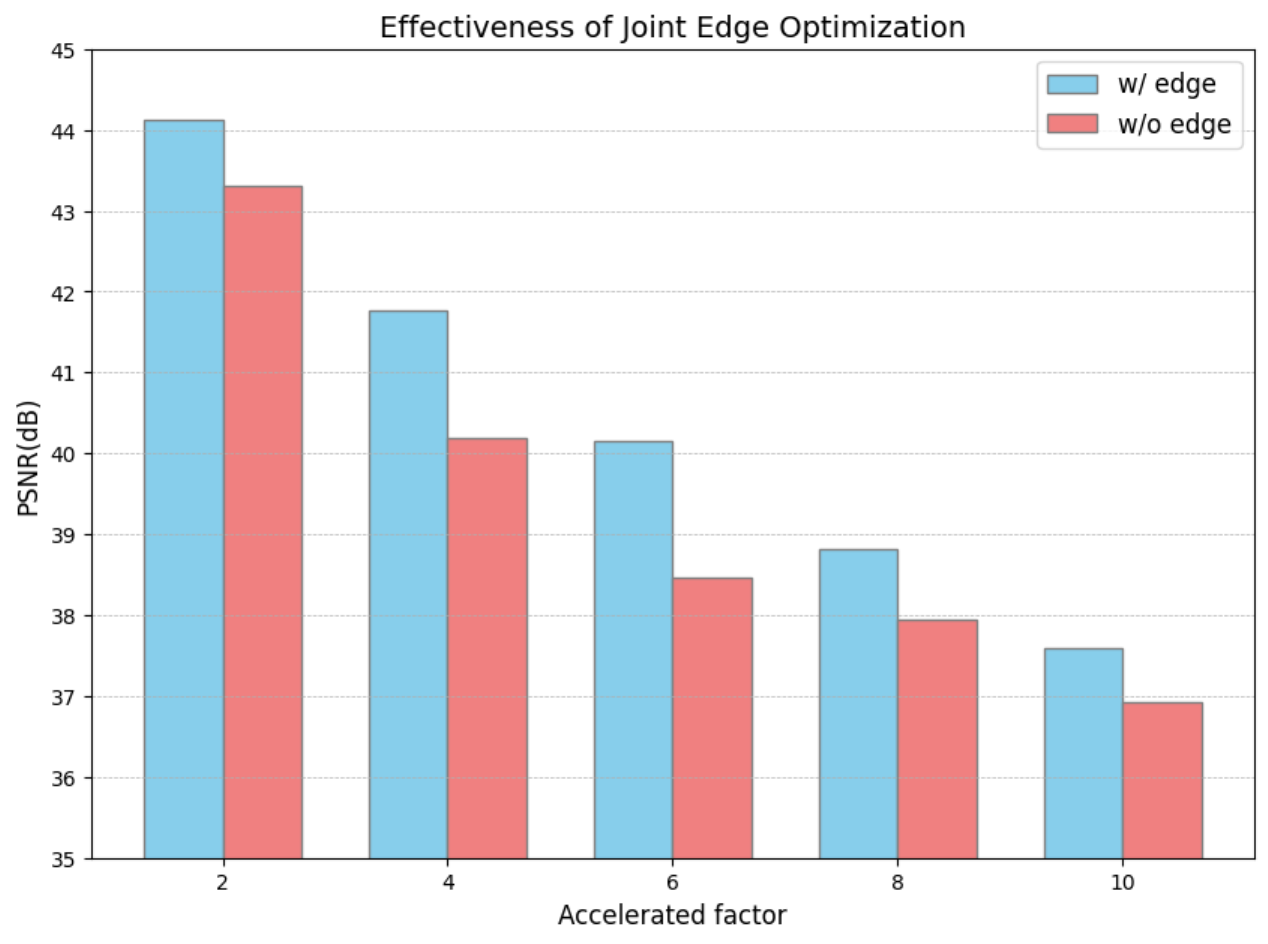} 
\caption{The reconstruction quality comparison with or without the joint optimization of edge.}
\label{fig_edgeAblaBar}
\end{figure}

To verify the effectiveness of the joint edge information optimization mechanism, 
we compare the proposed model with the classical reconstruction model (\ref{eq:1}) with no edge information involved.
Similar to the optimization of the proposed method, the model (\ref{eq:1}) is transformed by introducing auxiliary variables:
\begin{equation}
\mathop {\min }\limits_{x,Z} \frac{1}{2}\sum\limits_{i = 1}^{{n}} {\left\| {UF{S_i}x - {y_i}} \right\|_2^2}  + \frac{\beta }{2}\left\| {Z - x} \right\|_2^2 + {\lambda _1}\Phi(Z).
\label{eq:18}
\end{equation}
Afterward, the variables are separated to learn the optimization of auxiliary variables using an image denoising network. Moreover, the reconstructed image is updated using a single-step gradient descent with the following iterative format:
\begin{equation}
{Z^{(k)}} = IDN({x^{(k)}},{\theta_i ^{(k)}}),
\label{eq:19}
\end{equation}
\begin{equation}
\begin{aligned}
{x^{(k + 1)}} = {x^{(k)}} &- s[\sum\limits_{i = 1}^{{n}} {S_i^H{F^H}{U^H}(UF{S_i}{x^{(k)}} - {y_i})}\\  &- \beta ({Z^{(k)}} - {x^{(k)}})].
\end{aligned}
\label{eq:20}
\end{equation}
Aside from the implementation of edge information optimization, the experimental configuration for model (\ref{eq:18}) aligns with the proposed method. 
We conduct a compared experiments on a multi-coil dataset with various accelerated factor ranging from 2 to 10 using random undersampling template.
The data presented in Figure \ref{fig_edgeAblaBar} clearly demonstrates the superiority of incorporating edge information in the image reconstruction process. We observe that integrating edge optimization markedly improves performance as evidenced by the increasement of PSNR metric. This enhancement supports the argument for the joint edge optimization mechanism's beneficial role in improving image reconstruction quality.

\subsubsection{Effectiveness of Deep Unfolding Network}
The architecture of our proposed method employs a deep unfolding network strategy by integrating an \emph{ERN} module and an \emph{IDN} module. These two network modules simulate the proximal operators associated with undetermined regularization and parameters in the proposed joint edge optimization model. To validate the effectiveness of the two deep unfolding modules, we conduct an ablation study on \emph{ERN} and \emph{IDN} on the multi-coil dataset at 6x random sampling rates.
The results are tabulated in Table \ref{table_unfold} offering a comparison of reconstruction outcomes.
From the compared results, the model's performance enhancement was observed when deep unfolding modules are involved. When the \emph{ERN} is involved, the PSNR value attain the increment of $3.2$ dB, validating the advantage of unfolding the edge-related proximal operator for image reconstruction. In addition, a further improvement of $1.5$ dB in PSNR values is observed compared to solely relying on edge-related proximal operator unfolding, which shows the positive effect of the \emph{ERN} modules.
These increments in performance metrics confirm the efficacy of the deep unfolding framework. By capitalizing on the inherent representational capabilities of neural networks, our method excels in reconstructing MR images of superior quality.

\begin{table}[t]
\centering
\begin{tabular}{cc|cc}
\hline 
IDN & ERN & PSNR & SSIM \\
\hline 
\XSolidBrush & \XSolidBrush & 35.34 & 0.9278 \\
\XSolidBrush & \Checkmark & 38.61 & 0.9677\\
\Checkmark & \XSolidBrush & 39.45 & 0.9753\\
\Checkmark & \Checkmark & \textbf{40.15} & \textbf{0.9843}\\
\hline 
\end{tabular}
\caption{The PSNR and SSIM values under different configuration of deep unfolding method}
\label{table_unfold}
\end{table}

\begin{figure}[t]
\centering
\includegraphics[width=1.0\columnwidth]{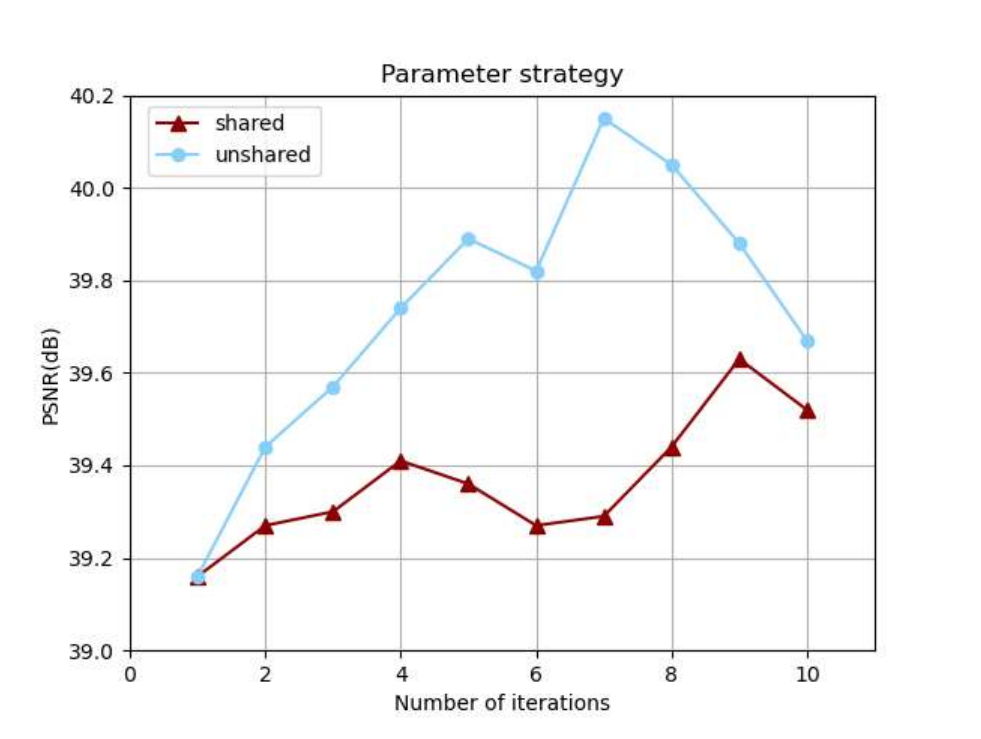} 
\caption{The reconstruction quality comparison line graph with two kinds of parameter strategy under different number of iteration.}
\label{fig_param_stg}
\end{figure}

\subsection{Discussion on the Iterative Parameter Strategy}

In a multi-stage deep unfolding reconstruction architecture, there are typically two parameter strategies available for selection: the recursive shared-parameter form and the non-shared form that assigns independent parameters to each iteration stage. The shared-parameter form is characterized by a compact parameter set, promoting speed up inference by minimizing the computational footprint. Conversely, the non-shared form delivers enhanced flexibility, with each stage fine-tuned for optimal performance at the potential cost of increased parameter count.

Our experimental involved testing the performance impact of both strategies across different iteration stages ranging from 1 to 10 on a multi-coil dataset subjected to 6x random sampling. The outcomes of this investigation are graphically presented in Figure \ref{fig_param_stg}, providing quantitative evidence for the comparative analysis.
The result curves illustrate a superior performance yield when deploying the non-shared parameter configuration within our proposed model. This suggests that the capability to customize each stage's parameters independently facilitates more effective optimization. Furthermore, an analysis of the trend in reconstruction quality as a function of iterative stages reveals a positive trajectory up until the seventh iteration in the non-shared configuration.
Additional iterations beyond this threshold will lead to a decrease in overall performance, a phenomenon we attribute to the increasing parameter scale that lead the model to be overfitting. 
These observations affirm the importance of tailoring the iterative parameter strategy to the specific requirements of the reconstruction task. Our proposed method, employing the non-shared parameter strategy, appears to reconcile the model's complexity with the imperative for high-quality image reconstruction.

\section{Conclusion}

In this paper, we propose a joint edge optimization deep unfolding network for accelerated MRI reconstruction. In order to explicitly and efficiently utilize the edge information, we design a novel edge utilization mechanism to explicitly optimize the edge information and use it to guide the reconstruction of high-quality MR images. Specifically, we correlate the edge information with the non-edge probability map and design a reconstruction model that incorporates edge-independent constraint terms along with co-regularizer terms between edge and MR image. In addition, we utilize the deep neural networks to unfold the optimization process of the reconstruction model aiming to automatically learn the prior information of MR images and edges. The ablation study verifies the effectiveness of the proposed edge utilization mechanism and the deep unfolding network modules.  In the quantitative and qualitative evaluation, our method outperforms the compared methods under different sampling schemes and sampling factors. In future work, we will explore the possibility of joint edge optimization deep unfolding networks for other low-level vision tasks.

\end{document}